\documentclass[12pt]{iopart}

\expandafter\let\csname equation*\endcsname\relax
\expandafter\let\csname endequation*\endcsname\relax

\usepackage{amsmath}
\usepackage{amssymb}

\begin{document}
\pagenumbering{roman}

\title{Revaluation of Mbelek and Lachi\`{e}ze-Rey scalar tensor theory of
gravitation to explain the measured forces in asymmetric resonant cavities}
\author{F. O. Minotti$^{1,2}$}

\begin{abstract}
The scalar-tensor theory of gravitation proposed by Mbelek and Lachi\`{e}
ze-Rey has been shown to lead to a possible explanation of the forces measured in asymmetric resonant microwave cavities \cite{minotti}. However, in the derivation of the equations from the action principle some inconsistencies were observed, like the need no to vary the electromagnetic invariant in a scalar source term. Also, the forces obtained were too high, in view of reconsideration of the experiments originally reported and of newly published results \cite{NASA}. In the present work the equations are re-derived using the full variation of the action, and also the constant of the theory re-evaluated employing the condition that no anomalous gravitational effects are produced by the earth's magnetic field. It is shown that the equations originally employed were correct, and that the newly evaluated constant gives the correct magnitude for the forces recently reported
\end{abstract}

\pagenumbering{roman}

\address{$^1$Universidad de Buenos Aires, Facultad de Ciencias Exactas y
Naturales, Departamento de F\'{\i}sica, Buenos Aires, Argentina.\\$^2$CONICET-Universidad de Buenos Aires, Instituto de F\'{\i}sica del Plasma (INFIP), Buenos Aires, Argentina}

\ead{minotti@df.uba.ar}
\maketitle

\section{Introduction}

The scalar-tensor (ST) gravitational theory of Mbelek and Lachi\`{e}ze-Rey
(MLR) \cite{MLR}, has the particularity of allowing electromagnetic (EM)
fields to modify the space-time metric far more strongly than predicted by
General Relativity and standard ST theories. The theory was applied in
cosmological \cite{mbelek2003} and galactic \cite{mbelek2004} contexts, and
in \cite{MLR} it was used to explain the discordant measurements of Newton
gravitational constant as due to the effect of the earth's magnetic field.
In \cite{minotti} it was further shown that a ST theory of the MLR type
could explain the unusual forces on asymmetric resonant cavities reported at
that time \cite{juan}. However, in the derivation of the equations from the
action principle some inconsistencies were observed, and also the forces
obtained, after reconsideration of the experiments originally reported and
new results \cite{NASA}, were too high. In the present work the equations
are re-derived using the full variation of the action, and the constant of
the theory re-evaluated with the consideration that no anomalous
gravitational effects are produced by the earth's magnetic field. It is
shown that the originally employed equations were correct, and that
employing the new evaluated constant a correct magnitude for the forces
reported recently is obtained.

\section{The MLR ST theory}

The details of the considered MLR ST theory were already presented in \cite%
{minotti} and \cite{raptis}, so that only a brief account will be given
here. The MRL action is given by (SI units are used) 
\begin{eqnarray}
S &=&-\frac{c^{3}}{16\pi G_{0}}\int \sqrt{-g}\phi Rd\Omega +\frac{c^{3}}{
16\pi G_{0}}\int \sqrt{-g}\frac{\omega \left( \phi \right) }{\phi }\nabla
^{\nu }\phi \nabla _{\nu }\phi d\Omega  \notag \\
&&+\frac{c^{3}}{16\pi G_{0}}\int \sqrt{-g}\phi \left[ \frac{1}{2}\nabla
^{\nu }\psi \nabla _{\nu }\psi -U\left( \psi \right) -J\psi \right] d\Omega 
\notag \\
&&-\frac{\varepsilon _{0}c}{4}\int \sqrt{-g}\lambda \left( \phi \right)
F_{\mu \nu }F^{\mu \nu }d\Omega -\frac{1}{c}\int \sqrt{-g}j^{\nu }A_{\nu
}d\Omega  \notag \\
&&+\frac{1}{c}\int \mathcal{L}_{mat}d\Omega .  \label{SKK}
\end{eqnarray}
In (\ref{SKK}) the internal, non-dimensional scalar field is $\phi $, while $%
\psi $ is an external scalar field introduced by MRL to stabilize the
reduced action of Kaluza-Klein multidimensional theories. These fields have
vacuum expectation values (VEV) $\phi _{0}$ and $\psi _{0}$, respectively. $%
G_{0}$ is Newton gravitational constant, with $G_{0}/\phi $ being its value
in vacuum (this means that $\phi _{0}=1$), $c$ is the velocity of light in
vacuum, and $\varepsilon _{0}$ is the vacuum permittivity. $\mathcal{L}%
_{mat} $ is the lagrangian density of matter. The other symbols are also
conventional, $R$ is the Ricci scalar, and $g$ the determinant of the metric
tensor $g_{\mu \nu }$. The Brans Dicke parameter $\omega \left( \phi \right) 
$ is considered a function of $\phi $. The function $\lambda \left( \phi
\right) $ in the term of the action of the EM field is found in reduced,
effective theories \cite{mbelek2003}. The EM tensor is $F_{\mu \nu }=\nabla
_{\mu }A_{\nu }-\nabla _{\nu }A_{\mu }$, \ given in terms of the EM
quadri-vector $A_{\nu }$, with sources given by the quadri-current $j^{\nu }$
. $U$ and $J$ are, respectively, the potential and source of the field $\psi 
$. The source $J$ contains contributions from the matter, EM field and the
scalar $\phi $. The model for $J$ proposed in \cite{MLR} is 
\begin{equation}
J=\beta _{mat}\left( \psi ,\phi \right) \frac{8\pi G_{0}}{c^{4}}%
T^{mat}+\beta _{EM}\left( \psi ,\phi \right) \frac{4\pi G_{0}\varepsilon
_{0} }{c^{2}}F_{\mu \nu }F^{\mu \nu },  \label{source}
\end{equation}
where $T^{mat}$ is the trace of the energy-momentum tensor of matter,\ 
\begin{equation*}
T_{\mu \nu }^{mat}=-\frac{2}{\sqrt{-g}}\frac{\delta \mathcal{L}_{mat}}{
\delta g^{\mu \nu }}.
\end{equation*}
The $\beta $ coefficients are unknown functions of the scalars, but in the
weak-field (WF) approximation they only contribute through the values of
their first-order derivatives at the VEV $\phi _{0}$ and $\psi _{0}$, and
thus appear as adjustable constants. Variation of (\ref{SKK}) with respect
to $g^{\mu \nu }$ results in ($T_{\mu \nu }^{EM}$ is the usual
electromagnetic energy tensor) 
\begin{eqnarray}
\phi \left( R_{\mu \nu }-\frac{1}{2}Rg_{\mu \nu }\right) &=&\frac{8\pi G_{0} 
}{c^{4}}\left[ \lambda \left( \phi \right) T_{\mu \nu }^{EM}+T_{\mu \nu
}^{mat}\right] +T_{\mu \nu }^{\phi }  \notag \\
&&+\frac{\phi }{2}\left( \nabla _{\mu }\psi \nabla _{\nu }\psi -\frac{1}{2}
\nabla ^{\gamma }\psi \nabla _{\gamma }\psi g_{\mu \nu }\right)  \notag \\
&&+\frac{\phi }{2}\left( U+J\psi \right) g_{\mu \nu }-\phi \psi \frac{\delta
J}{\delta g^{\mu \nu }},  \label{Glm}
\end{eqnarray}
where $T_{\mu \nu }^{\phi }$ is the energy tensor of the scalar $\phi $ 
\begin{equation*}
T_{\mu \nu }^{\phi }=\nabla _{\mu }\nabla _{\nu }\phi -\nabla ^{\gamma
}\nabla _{\gamma }\phi g_{\mu \nu }+\frac{\omega \left( \phi \right) }{\phi }
\left( \nabla _{\mu }\phi \nabla _{\nu }\phi -\frac{1}{2}\nabla ^{\gamma
}\phi \nabla _{\gamma }\phi g_{\mu \nu }\right) ,
\end{equation*}%
and where, since $\delta T^{mat}/\delta g^{\mu \nu }=0$, 
\begin{equation*}
\frac{\delta J}{\delta g^{\mu \nu }}=\frac{4\pi G_{0}\varepsilon _{0}}{c^{2}}
\beta _{EM}\left( \psi ,\phi \right) F_{\mu \theta }F_{\nu \xi }g^{\theta
\xi }.
\end{equation*}
Variation with respect to $\phi $ gives 
\begin{eqnarray*}
\phi R+2\omega \nabla ^{\nu }\nabla _{\nu }\phi &=&\left( \frac{\omega }{
\phi }-\frac{d\omega }{d\phi }\right) \nabla ^{\nu }\phi \nabla _{\nu }\phi
- \frac{4\pi G_{0}\varepsilon _{0}}{c^{2}}\phi \frac{d\lambda }{d\phi }
F_{\mu \nu }F^{\mu \nu } \\
&&-\frac{\partial J}{\partial \phi }\psi \phi +\phi \left[ \frac{1}{2}\nabla
^{\nu }\psi \nabla _{\nu }\psi -U\left( \psi \right) -J\psi \phi \right] ,
\end{eqnarray*}
which can be rewritten, using the contraction of (\ref{Glm}) with $g^{\mu
\nu }$ to replace $R$, as 
\begin{eqnarray}
\left( 2\omega +3\right) \nabla ^{\nu }\nabla _{\nu }\phi &=&-\frac{d\omega 
}{d\phi }\nabla ^{\nu }\phi \nabla _{\nu }\phi -\frac{4\pi G_{0}\varepsilon
_{0}}{c^{2}}\phi \frac{d\lambda }{d\phi }F_{\mu \nu }F^{\mu \nu }  \notag \\
&&+\frac{8\pi G_{0}}{c^{4}}T^{mat}\left( 1+2\psi \phi \beta _{mat}\right)
+\phi U\left( \psi \right)  \notag \\
&&-\left( J+\frac{\partial J}{\partial \phi }\phi \right) \psi \phi ,
\label{phi}
\end{eqnarray}
where it was used that $T^{EM}=T_{\mu \nu }^{EM}g^{\mu \nu }=0$. The
non-homogeneous Maxwell equations are obtained by varying (\ref{SKK}) with
respect to $A_{\nu }$, 
\begin{equation}
\nabla _{\mu }\left\{ \left[ \lambda \left( \phi \right) +\phi \psi \beta
_{EM}\left( \psi ,\phi \right) \right] F^{\mu \nu }\right\} =\mu _{0}j^{\nu
}.  \label{Maxwell}
\end{equation}
with $\mu _{0}$ the vacuum permeability. Finally, the variation with respect
to $\psi $ results in 
\begin{equation}
\nabla ^{\nu }\nabla _{\nu }\psi +\frac{1}{\phi }\nabla ^{\nu }\psi \nabla
_{\nu }\phi =-\frac{\partial U}{\partial \psi }-\left( J+\frac{\partial J}{
\partial \psi }\psi \right) .  \label{psi}
\end{equation}

\section{Analysis}

A crucial point is indicated by Maxwell equations (\ref{Maxwell}) because it
turns out that the expected magnitude of $\phi \psi \partial \beta
_{EM}\left( \psi ,\phi \right) /\partial \phi $ at the VEV of the scalar
fields is very large, which would clearly result in the wrong Maxwell
equations. First, we note that, in order to recover Einstein-Maxwell
equations when the scalar fields are not excited, the potential $U\left(
\psi \right) $ and its derivative are zero when evaluated at the VEV $\psi
_{0}$, $\lambda \left( \phi _{0}\right) =1$, and 
\begin{equation}
\beta _{mat}\left( \psi _{0},\phi _{0}\right) =\beta _{EM}\left( \psi
_{0},\phi _{0}\right) =0.  \label{betasvev}
\end{equation}%
One can then readily conclude that when the scalar fields are close to their
VEV the correct Maxwell equations are obtained only if 
\begin{equation}
\nabla _{\mu }\psi =-\frac{\partial \left( \psi \phi \beta _{EM}\right)
/\partial \phi }{\partial \left( \psi \phi \beta _{EM}\right) /\partial \psi 
}\nabla _{\mu }\phi \equiv h\left( \psi ,\phi \right) \nabla _{\mu }\phi .
\label{hcondition}
\end{equation}%
In this relation we have left out the function $\lambda \left( \phi
\right) $, which is responsible for possible effects of the scalar
field on the electromagnetic fields \cite{raptis}. Employing relation (\ref
{hcondition}) in Eq. (\ref{psi}) we obtain 
\begin{eqnarray}
\nabla ^{\nu }\nabla _{\nu }\phi  &=&-\left( \frac{1}{h}\frac{\partial h}{%
\partial \phi }+\frac{\partial h}{\partial \psi }+\frac{1}{\phi }\right)
\nabla ^{\nu }\phi \nabla _{\nu }\phi   \notag \\
&&-\frac{1}{h}\left( \frac{\partial U}{\partial \psi }+J+\psi \frac{\partial
J}{\partial \psi }\right) ,  \label{phicondition}
\end{eqnarray}%
which is to be compared with Eq. (\ref{phi}). The first thing to note is
that this implies that 
\begin{equation}
\frac{1}{h}\frac{\partial h}{\partial \phi }+\frac{\partial h}{\partial \psi 
}+\frac{1}{\phi }=\frac{d\omega /d\phi }{2\omega +3}.  \label{wps2wp3}
\end{equation}%
Since the right-hand side is a function of only $\phi $, for reasons to be
made clear immediately, it is expected that at the scalars VEV 
\begin{equation*}
\frac{1}{h}\frac{\partial h}{\partial \phi }+\frac{\partial h}{\partial \psi 
}=0,
\end{equation*}%
which, employing the definition of $h$ in (\ref{hcondition}), is written as%
\begin{equation}
2h\frac{\partial ^{2}\left( \psi \phi \beta _{EM}\right) }{\partial \phi
\partial \psi }+\frac{\partial ^{2}\left( \psi \phi \beta _{EM}\right) }{%
\partial \phi ^{2}}+h^{2}\frac{\partial ^{2}\left( \psi \phi \beta
_{EM}\right) }{\partial \psi ^{2}}=0.  \label{betasd2}
\end{equation}%
With the condition (\ref{betasvev}), relation (\ref{betasd2}) is satisfied
at the scalars VEV if all second derivatives of $\beta _{EM}$ are zero at
those scalar values. This is a self-consistent condition that also means
that the equations around these values can be determined at second order in
the perturbed scalars with the only parameter of the first derivatives of $%
\beta _{EM}$ at $\left( \psi _{0},\phi _{0}\right) $. In this way we obtain
from (\ref{wps2wp3}) 
\begin{equation}
\frac{\omega _{0}^{\prime }}{2\omega _{0}+3}=1,  \label{omegarelat}
\end{equation}%
where $\omega _{0}\equiv \omega \left( \phi _{0}\right) $, $\omega
_{0}^{\prime }\equiv \left( d\omega /d\phi \right) _{\phi _{0}}$. Also, from
the comparison between (\ref{phi}) and (\ref{phicondition}), 
\begin{subequations}
\label{dbetas}
\begin{eqnarray}
\frac{\psi _{0}}{h_{0}}\left. \frac{\partial \beta _{mat}}{\partial \psi }%
\right\vert _{\left( \psi _{0},\phi _{0}\right) } &=&\frac{1}{2\omega _{0}+3}%
\left( \psi _{0}\left. \frac{\partial \beta _{mat}}{\partial \phi }%
\right\vert _{\left( \psi _{0},\phi _{0}\right) }-1\right) , \\
\frac{\psi _{0}}{h_{0}}\left. \frac{\partial \beta _{EM}}{\partial \psi }%
\right\vert _{\left( \psi _{0},\phi _{0}\right) } &=&\frac{1}{2\omega _{0}+3}%
\left( \psi _{0}\left. \frac{\partial \beta _{EM}}{\partial \phi }%
\right\vert _{\left( \psi _{0},\phi _{0}\right) }+\lambda _{0}^{\prime
}\right) ,
\end{eqnarray}%
where $\lambda _{0}^{\prime }\equiv \left( d\lambda /d\phi \right) _{\phi
_{0}}$, and 
\end{subequations}
\begin{equation*}
h_{0}=-\left. \frac{\partial \beta _{EM}/\partial \phi }{\partial \beta
_{EM}/\partial \psi }\right\vert _{\left( \psi _{0},\phi _{0}\right) }.
\end{equation*}
With the assumption that the dominant term in the bracket on the
right-hand side of the second of (\ref{dbetas}) is the first one (the
consistency is checked below), one obtains 
\begin{equation}
\left. \frac{\partial \beta _{EM}}{\partial \psi }\right\vert _{\left( \psi
_{0},\phi _{0}\right) }=\pm \frac{1}{\left\vert 2\omega _{0}+3\right\vert
^{1/2}}\left. \frac{\partial \beta _{EM}}{\partial \phi }\right\vert
_{\left( \psi _{0},\phi _{0}\right) },  \label{h0det}
\end{equation}
with the additional condition that $2\omega _{0}+3<0$. As is well
known, this condition would lead to an unstable action for a pure
Brans-Dicke theory. However, note that from (\ref{h0det}) one has $
h_{0}=\mp \left\vert 2\omega _{0}+3\right\vert ^{1/2}$, which,
together with relation (\ref{hcondition}), means that the MRL action (\ref
{SKK}) has, "on-shell", and at least around the scalars VEV, an effective
positive Brans-Dicke parameter, which solves the instability problem, and
what motivated the introduction of the external escalar field in MRL theory.

\section{Weak-field approximation}

With the considerations given above, Eq. (\ref{Glm}) can be approximated in
the WF limit keeping only the lowest significant order in the perturbations $%
h_{\mu \nu }$ of the metric $g_{\mu \nu }$ about the Minkowski metric $\eta
_{\mu \nu }$, with signature (1,-1,-1,-1), and of the scalar fields about
their VEV $\phi _{0}$ and $\psi _{0}$, as 
\begin{equation}
-\eta ^{\gamma \delta }\partial _{\gamma \delta }\overline{h}_{\mu \nu
}=2\left( \partial _{\mu \nu }\phi -\eta ^{\gamma \delta }\partial _{\gamma
\delta }\phi \eta _{\mu \nu }\right) ,  \label{Gik0}
\end{equation}%
with the Lorentz gauge 
\begin{equation}
\partial _{\gamma }\overline{h}_{\nu }^{\gamma }=0,  \label{LG}
\end{equation}%
where 
\begin{equation*}
\overline{h}_{\mu \nu }\equiv h_{\mu \nu }-\frac{1}{2}h\eta _{\mu \nu },
\end{equation*}%
with $\overline{h}=\eta ^{\mu \nu }\overline{h}_{\mu \nu }$. The WF
equations for the scalars considering only the electromagnetic sources are
then 
\begin{equation}
\eta ^{\gamma \delta }\partial _{\gamma \delta }\phi =-\Gamma F_{\mu \nu
}F^{\mu \nu },  \label{phiWF}
\end{equation}%
and%
\begin{equation*}
\eta ^{\gamma \delta }\partial _{\gamma \delta }\psi =\pm \left\vert 2\omega
_{0}+3\right\vert ^{1/2}\Gamma F_{\mu \nu }F^{\mu \nu },
\end{equation*}
with 
\begin{equation}
\Gamma =\frac{4\pi G_{0}\varepsilon _{0}}{c^{2}}\frac{\psi _{0}}{2\omega
_{0}+3}\left. \frac{\partial \beta _{EM}}{\partial \phi }\right\vert
_{\left( \psi _{0},\phi _{0}\right) }.  \label{Gamon}
\end{equation}
As can be seen in \cite{minotti}, for slow moving neutral masses, the WF
limit of the geodesic equation corresponds to motion in flat Minkowski
space-time under the action of a specific force (per unit mass) given by
(Latin indices correspond to the spatial coordinates) 
\begin{equation}
f_{i}=-\frac{c^{2}}{4}\frac{\partial }{\partial x_{i}}\left( \overline{h}%
_{00}+\overline{h}_{kk}\right) +c\frac{\partial \overline{h}_{0i}}{\partial
t }.  \label{forcepermass}
\end{equation}
Using Eqs. (\ref{Gik0}) and (\ref{forcepermass}) the gravitational field is
thus represented in the WF approximation by a gravitational potential $\chi $
in flat Minkowski space-time, whose determining equation with only scalar
field sources is ($\square \equiv \eta ^{\gamma \delta }\partial _{\gamma
\delta }$) 
\begin{eqnarray}
\square \chi &=&\frac{\partial ^{2}\phi }{\partial t^{2}}-\frac{c^{2}}{2}
\square \phi  \notag \\
&=&\frac{\partial ^{2}\phi }{\partial t^{2}}+\frac{c^{2}\Gamma }{2}F_{\mu
\nu }F^{\mu \nu },  \label{kapafinal}
\end{eqnarray}%
where (\ref{phiWF}) was used to write the second line.

\section{Constant determination}

As was argued in \cite{minotti}, in order for the MLR theory to be
consistent with the lack of strong gravitational effects due to the magnetic
field of the Earth, the non-linear terms in Eq. (\ref{phi}) should be kept,
even in the WF approximation. This is so because the laplacian terms in
these equations can be zero for this particular type of source, and thus the
equalities are satisfied by the higher order terms. Specifically, for a
static magnetic field outside its sources one can write $\mathbf{B}=\mathbf{%
\nabla }\Psi $, with $\nabla ^{2}\Psi =0$, so that, from Eq. (\ref{phi}),
one has (here we use the consistency condition that the second derivatives
of $\beta _{EM}\left( \psi ,\phi \right) $ are zero at the VEV values) 
\begin{equation}
\nabla ^{2}\phi +\frac{\omega _{0}^{\prime }}{2\omega _{0}+3}\mathbf{\nabla }%
\phi \cdot \mathbf{\nabla }\phi =2\Gamma B^{2},  \label{phiearthb}
\end{equation}%
Eq. (\ref{phiearthb}) has the solution, using also relation (\ref{omegarelat}
), 
\begin{equation}
\phi =1+\sqrt{2\Gamma }\Psi .  \label{phiearth}
\end{equation}%
Noting that the measured gravitational constant is given by $G=G_{0}/\phi $,
one can repeat the analysis made in \cite{MLR} using expression (\ref%
{phiearth}) to obtain the values of $G_{0}$ and $\Gamma $ that gives the
best fit between the locally measured value of $G$ and $G_{0}/\phi $, with $%
\phi $ given by (\ref{phiearth}) with the local value of the earth's
magnetic field given in terms of $\Psi $ expressed, using Gauss coeffcients,
by \cite{MLR} 
\begin{equation}
\Psi =\frac{a^{3}}{r^{2}}\left( g_{1}^{0}\cos \theta +g_{1}^{1}\sin \theta
\cos \varphi +h_{1}^{1}\sin \theta \sin \varphi \right) ,  \label{bigpsi}
\end{equation}%
where $\varphi $ is the longitude, $\theta $ the co-latitude, $r$ the
distance to the earth center, and $a$ the earth radius. Employing the same
database and parameters in (\ref{bigpsi}) as in \cite{MLR} we obtain (SI
units) 
\begin{eqnarray*}
G_{0} &=&6.6752\times 10^{-11}\,\text{Nm}^{2}\text{kg}^{-2}, \\
\Gamma  &=&3.35\times 10^{-12}\,\text{A}^{2}\text{N}^{-2}.
\end{eqnarray*}%
Note that, from (\ref{Gamon}), with this value of $\Gamma $, $\phi \psi
\partial \beta _{EM}/\partial \phi $ at the scalars VEV has a very large
value, as assumed. It is worth noting that this evaluation of $
\Gamma $ is not completetly satisfactory because there are not
enough data and with sufficient spread of the magnetic
field values at the sites where $G$ was measured to have a
reliable statistics. It is made here only to check that the theory can be
consistent with different experimental results.

\section{Force evaluation}

In this section we present the basis for the evaluation of the force on
axially symmetric cavities, including an expression more convenient than
that used in \cite{minotti} because the involved derivatives can be
performed explicitly so that the final expression reduces to a
(five-dimensional) integral. From what was seen in the previous section, in
the weak field approximation, the average thrust is determined as a
gravitational force derived from the gravitational potential $\chi $, whose
determining equation is obtained from Eq. (\ref{kapafinal}) by taking the
time average 
\begin{equation}
\nabla ^{2}\chi =-\varkappa \Sigma ,  \label{chi}
\end{equation}%
where the constant $\varkappa $ has the value (SI units) 
\begin{equation*}
\varkappa \equiv c^{2}\Gamma =3.0\times 10^{5}\,\text{A}^{2}\text{s}^{2}%
\text{kg}^{-2},
\end{equation*}
and the source $\Sigma $ is the time average of\ the field invariant $%
B^{2}-E^{2}/c^{2}$ 
\begin{equation}
\Sigma =\left\langle B^{2}-E^{2}/c^{2}\right\rangle .  \label{sourceS}
\end{equation}
The obtained value of $\varkappa$ is a factor $10^{-4}$ of that used in \cite%
{minotti}. In this way, if (\ref{chi}) is employed to determine the force on
the asymmetric resonant cavity employed in \cite{NASA}, one obtains forces
of the order of $\mu $N/W, as those reported. If $\rho $ is the mass density
of the material of the cavity the force on a cavity with thin walls is then
given by 
\begin{equation}
\mathbf{F}=-\rho \int \lambda \mathbf{\nabla }\chi dS,  \label{thrust}
\end{equation}
where $\lambda $ is the local thickness of the cavity wall, and the surface
integral is extended to the whole surface of the cavity. If the cavity and
field distribution have axial symmetry, using spherical coordinates with the
angle $\theta $ measured from the cavity axis, the solution of (\ref{chi})
can be written as 
\begin{eqnarray}
\chi \left( r,\theta \right) &=&\frac{\varkappa }{\pi }\int \frac{\Sigma
\left( r^{\prime },\theta ^{\prime }\right) }{\sqrt{r^{2}+r^{\prime
2}-2rr^{\prime }\cos \left( \theta ^{\prime }-\theta \right) }}  \notag \\
&&\times K\left( -\frac{4rr^{\prime }\sin \theta \sin \theta ^{\prime }}{
r^{2}+r^{\prime 2}-2rr^{\prime }\cos \left( \theta ^{\prime }-\theta \right) 
}\right) r^{\prime 2}\sin \theta ^{\prime }dr^{\prime }d\theta ^{\prime }
\label{final}
\end{eqnarray}
where $K$ is the complete elliptic integral of the first kind, and the
volume integral is extended to the interior of the cavity. For the same
axial symmetry, the force (\ref{thrust}) has component only along the
direction of the cavity axis, which we call $z$, and is given by 
\begin{equation}
F_{z}=-\varrho \int \lambda \left( r,\theta \right) \left( \cos \theta \frac{
\partial \chi }{\partial r}-\frac{\sin \theta }{r}\frac{\partial \chi }{
\partial \theta }\right) dS.  \label{fz}
\end{equation}
Using expression (\ref{final}) in (\ref{fz}) the derivatives can be
explicitly made and one finally has 
\begin{eqnarray}
F_{z} &=&-\frac{\varkappa \rho }{\pi }\int \frac{\lambda \left( r,\theta
\right) \Sigma \left( r^{\prime },\theta ^{\prime }\right) \left( r^{\prime
}\cos \theta ^{\prime }-r\cos \theta \right) }{\left[ r^{2}+r^{\prime
2}-2rr^{\prime }\cos \left( \theta ^{\prime }+\theta \right) \right] \sqrt{
r^{2}+r^{\prime 2}-2rr^{\prime }\cos \left( \theta ^{\prime }-\theta \right) 
}}  \notag \\
&&\times E\left( -\frac{4rr^{\prime }\sin \theta \sin \theta ^{\prime }}{
r^{2}+r^{\prime 2}-2rr^{\prime }\cos \left( \theta ^{\prime }-\theta \right) 
}\right) r^{\prime 2}\sin \theta ^{\prime }dr^{\prime }d\theta ^{\prime }dS
\label{force_double}
\end{eqnarray}
where $E$ is the complete elliptic integral of the second kind. The
integration over the primed variables is extended to the volume of the
cavity, while the non-primed variables are integrated over the surface of
the cavity. If a single electromagnetic mode, axially symmetric is
considered, with angular frequency $\omega $, the field invariant $%
B^{2}-E^{2}/c^{2}$ can be written in general as 
\begin{equation*}
B^{2}-E^{2}/c^{2}=F_{B}\left( r,\theta \right) \cos ^{2}\left( \omega
t\right) -F_{E}\left( r,\theta \right) \sin ^{2}\left( \omega t\right),
\end{equation*}
and so its time average is given simply by 
\begin{equation*}
\Sigma=\left\langle B^{2}-E^{2}/c^{2}\right\rangle =\frac{1}{2}\left(
F_{B}-F_{E}\right).
\end{equation*}
As done in \cite{minotti} the cavity is approximated by a cone whose axis is
taken as the $z$ direction. The lateral wall corresponds to the spherical
angle $\theta =\theta _{0}$ (half angle of the cone), and the spherical caps
to the radii $r=r_{1,2}$, with $r_{2}>r_{1}$. The resonant modes are
standing electromagnetic waves satisfying the vector wave equation ($\mathbf{%
F}$ stands for either the electric field $\mathbf{E}$ , or the magnetic
induction $\mathbf{B}$) 
\begin{equation*}
\frac{1}{c^{2}}\frac{\partial ^{2}\mathbf{F}}{\partial t^{2}}-\nabla ^{2} 
\mathbf{F}=0.
\end{equation*}
The modes with axial symmetry and $\mathbf{B}$ transverse to the $z$
direction $\mathbf{e}_{z}$ (called the TM modes) that satisfy this equation
are (spherical coordinates are employed, with unit vectors $\mathbf{e}_{r}$, 
$\mathbf{e}_{\theta }$\ and $\mathbf{e}_{\varphi }$) 
\begin{eqnarray}
\mathbf{B} &=&-CkR\left( r\right) Q^{\prime }\left( \theta \right) \cos
\left( \omega t\right) \mathbf{e}_{\varphi },  \label{bmode} \\
\mathbf{E}/c &=&C\left\{ \frac{R\left( r\right) }{r}n\left( n+1\right)
Q\left( \theta \right) \mathbf{e}_{r}\right.  \notag \\
&&\left. +\left[ \frac{R\left( r\right) }{r}+R^{\prime }\left( r\right) %
\right] Q^{\prime }\left( \theta \right) \mathbf{e}_{\theta }\right\} \sin
\left( \omega t\right)  \label{emode}
\end{eqnarray}
where $C$ is a global constant. The complementary set of modes with $\mathbf{%
E}$ transverse to the $z$ direction (TE modes) have the expressions (\ref%
{bmode}) and (\ref{emode}), but with $\mathbf{E}/c$ and $\mathbf{B}$
interchanged. The functions $R$ and $Q$ are defined as 
\begin{eqnarray*}
Q\left( \theta \right) &=&P_{n}\left( \cos \theta \right) , \\
R\left( r\right) &=&R_{+}\left( r\right) \cos \alpha +R_{-}\left( r\right)
\sin \alpha , \\
R_{\pm }\left( r\right) &=&\frac{J_{\pm \left( n+1/2\right) }\left(
kr\right) }{\sqrt{r}},
\end{eqnarray*}
where $P_{n}$ is the Legendre polynomial of order $n$, $J_{m}$ the Bessel
function of the first kind of order $m$. The constants $\alpha $ and $k$ are
determined along with the order $n$ by imposing the boundary conditions of
zero normal component of the magnetic field and of zero tangential
components of the electric field at the metallic walls. Finally, the quality
factor of the cavity, $Q_{cav}$, for each mode is conventionally defined as 
\begin{equation}
Q_{cav}\equiv \frac{\omega \left\langle U\right\rangle }{\left\langle
W\right\rangle },  \label{quw}
\end{equation}
where $\omega =kc$ is the angular frequency of the mode, $\left\langle
U\right\rangle $ is the temporal average of its electromagnetic energy, and $%
\left\langle W\right\rangle $ is the average dissipated power in the wall
cavities. If the cavity is fed with an average electromagnetic power $P$, in
the permanent regime one has $\left\langle W\right\rangle =P$, and so, 
\begin{equation}
\left\langle U\right\rangle =\frac{\int \left\langle B^{2}\right\rangle dV}{
\mu _{0}}=\frac{Q_{cav}P}{\omega },  \label{csquare}
\end{equation}
which allows to determine the global constant $C$, given the fed average
power and the characteristics of the cavity for the considered mode.

The electromagnetic mode reported in \cite{NASA} is the TM212, which has no
axial symmetry, and thus the corresponding force cannot be simulated with ( %
\ref{force_double}) (the reported value is 1.2 $\mu$N/W, directed toward the
small end of the cone). For this reason we will instead present the results
of the model for a few axially symmetric TM and TE modes.

The cavity employed in \cite{NASA} is simulated here with corresponding caps
of radius 31.22 cm and 54.69 cm, and side walls inclined 14.8 degrees from
the axial direction. The copper side walls have 0.6 mm thickness, while the
caps are made of a thin copper layer (35.56 $\mu$m) over a 1.6 mm thick PCB
board.

The TM010 mode evaluated for a cavity of these dimensions has a frequency of
973.2 MHz with a quality factor of 34,000. The corresponding force evaluated
with (\ref{force_double}) has a magnitude of 0.69 $\mu$N/W and directed
toward the small end of the cone. For the TM012 mode the frequency is 1,6856
MHz and the quality factor 33,700. The force has a magnitude of 0.14 $\mu$%
N/W, directed toward the large end of the cone. For the TE010 mode the
frequency is 1,7309 MHz and the quality factor 70,800. The force has a
magnitude of 0.11 $\mu$N/W, directed toward the large end of the cavity. For
the TE012 mode the frequency is 2,1322 MHz and the quality factor 79,000.
The force has a magnitude of 0.011 $\mu$N/W, directed toward the large end
of the cavity.

From all these cases, only the TM010 mode has the same direction and a
magnitude comparable (about 57$\%$) to the force reported for the TM212
mode. The rest of the cases have a force directed toward the large end of
the cone, and magnitudes about 10$\%$ of those reported, with the exception
of the TE012 mode that has a very low value, only 1$\%$. Cases in which the
force changes direction and/or has very low magnitudes were previously
reported, but not so thoroughly studied as the published case.

\section{Conclusions}

We have rederived the equations of MRL theory employing a full variation of
the action, and estimated the constant of the WF limit of the theory,
consistent with the lack of anomalous effects by the earth's magnetic field. 
In this respect it is important to ascertain that expression (\ref
{phiearth}) is the pertinent solution of the non-linear equation (\ref
{phiearthb}). From this equation one can determine, for instance, that if
the magnetic field does not diverge the normal derivative of $\phi $
 across any surface must be continuous. Since also the normal component of
the magnetic field must be continuous across any surface, the solution (\ref
{phiearth}) satisfies the correct boundary conditions if the sources of the
magnetic field are confined to relatively thin sheets of electric current.
Also, one can argue that, of all possible solutions compatible with given
boundary conditions, those with minimum energy (defined in some appropriate
sense) should be those found in a real system. In this way it is expected
that the solution that minimizes the gravitational forces (with their
feedback effects on the material sources, for instance forcing the electric
current to be distributed on thin sheets) is the one verified in practice.
In this respect MRL theory can be of fundamental importance to our
understanding of the generation of geophysical and astrophysical magnetic
fields. Of course, an evaluation of the constant in a specifically designed and well controlled experiment should be desirable, the only purpose of the present estimation is to show the compatibility of the theory with all available related data

With all this, it can now be ascertained with more confidence that the theory is consistent and compatible with the reported effects. 

\end{document}